\documentclass[aps,preprint]{revtex4}%
\usepackage{amsfonts}
\usepackage{amsmath}
\usepackage{amssymb}
\usepackage{graphicx}%
\providecommand{\U}[1]{\protect\rule{.1in}{.1in}}

\begin{document}
\title{{\LARGE REALIZATION OF EINSTEIN
\'{}%
S MACHIAN PROGRAM }\\\textbf{ }}
\author{Marcelo Samuel Berman$^{1}$}
\affiliation{$^{1}$Instituto Albert Einstein/Latinamerica\ - }
\affiliation{Av. Sete de Setembro, 4500 - \ \# 101}
\affiliation{80250-210 - Curitiba - PR - Brazil }
\affiliation{emails: msberman@institutoalberteinstein.org , \ marsambe@yahoo.com, }
\affiliation{}
\date{Original Version: 03 SEPTEMBER 2011}

\begin{abstract}
\qquad\ \ \ \ \ \ \ \ \ \ \ \ \ \ The realization of Einstein
\'{}%
s Machian Program, is here accomplished.

\end{abstract}
\maketitle

\begin{center}

{\LARGE REALIZATION OF EINSTEIN
\'{}%
S MACHIAN PROGRAM }\textbf{ }

\bigskip Marcelo Samuel Berman

\end{center}

\bigskip{\LARGE 1. Introduction}

\bigskip The purpose of the present Section is to give a viewpoint of the
Einstein's Machian Program, hinting that, under these optics, it may have been
already attained through the possible rotation of the Universe.The Godlowski
method of introducing rotation in Cosmology will be presented, then we solved
the three NASA anomalies, which are the linear deceleration of the Pioneers in
outer space, the spin-down of the same space-craft, and the fly-by anomaly
that accompanies gravity assists, with a surplus of kinetic energy in the end.
In the previous book (Berman, 2011d), neither fly-by anomaly, nor the
Einstein's program, nor the different kinds of relativistic rotations had been
considered. According to Ni(2011), the ultra-precise Gravity Probe B
experiment analysis leaves open the cosmic polarization rotation (which may be
due to a Universal rotational state), and the limit of angular speeds attained
by this experiment can be checked from the abstract of his paper to be around
$10^{-17}s^{-1}$.

Though Einstein originally recognized Machian ideas as important, Barbour
(1990) describes that Einstein concentrated first on the construction of a
local gravitational theory, delaying consideration on the relativity of motion
to a future global approach. According to Einstein, one should not speak in a
gravitational theory, on absolute accelerations of a coordinate system, as
much as, in the Special Relativity Theory; one could not work with absolute
speeds of an observer. This apparent betrayal of Machian ideas was necessary
in order to create the field equations of General Relativity. Mach, on the
other side, placed the distinguished accelerated reference frame, within the
distant stars; i.e., the local distinguished reference frames could be
identified by looking at the Cosmos as a whole.

Godlowski (2011) has reviewed the universal rotational evidence. Fine-tuning
arguments can also be invoked in favor of such rotation. Gamow (1946)
considered that a rotation and expansion of the Universe could have the same
physical origin, and we equated the angular speed with the Hubble's parameter.
Ni (2008; 2009) shows that a rotation of 0.1 radians exists in the
polarization of CMBR. If we divide this angle by the age of the Universe, we
find about $10^{-19}$rad/s. Chechin (2010) finds the same result, by other
token. The present author, thinks that it is the ripe time now to reconsider
the role of the distant stars reference frame as a paradigm of accelerations,
and introduce the Universal rotation as proposed by Berman (2007). The angular
speed $\omega\cong\frac{c}{R}$ was adopted, because it carries a peculiar
rotational state of the Universe, with deceleration $a_{cp}=-\omega^{2}R$\,
which coincides with the Pioneers anomalous deceleration described in next
Section. In 2010, Berman and Gomide (2010) succeeded in making a full General
Relativistic model of the rotating and expanding Machian Universe that was
further generalized by Berman and Gomide (2011a) into a full class of models,
thus explaining the three NASA anomalies.

The proof that Gamow (1946) is correct, lies on the following calculation:
consider a Machian Universe (a ``spherical'' ball) and let us calculate the
Newtonian gravitational acceleration on the surface, $a_{g}=-\frac{GM}{R^{2}}
$ . If the zero-energy Universe is considered, so that the sum of the inertial
energy $Mc^{2}$ with the negative self-energy $\frac{GM^{2}}{R}$, and is
equated to zero, we find $\frac{GM}{c^{2}R}\cong1$. With this approximation,
we obtain $a_{g}\cong-\frac{c^{2}}{R}=-9\cdot10^{-8}\,\text{cm}/s^{2}=$ first
Pioneer anomaly.

On the other hand, when Godlovski's \textit{et al.} (2004) idea is taken into
account for a rotating Universe (see Section 3), we will find, neglecting the
cosmological constant, $\kappa\rho=6H^{2}$, and with $\rho=\frac{M}{\frac
{4}{3}\pi R^{3}}$, and $H\cong\omega$, we recover the result $\omega\cong%
\frac{c}{R}$, thus winding up with Berman's angular speed of the Universe. So
Einstein's Machian program now has the prototype of a distant stars
acceleration -- it is the Pioneers first anomaly in action, as we shall see next.

\bigskip

\bigskip{\LARGE 2. THE THREE NASA ANOMALIES}

\section{}

Several authors have alerted the scientific community about the fly-by
anomaly: during Earth gravity assists, spacecraft has suffered from an
extra-energization, characterized by a positive extra speed, such that,
measured ``at infinity'' the hyperbolic orbiting object presented an
empirically calculated $\Delta V/V$ around $10^{-6}.$ A formula was supplied,
(Anderson \textit{et al.}, 2008),
\[
\frac{\Delta V}{V}=\frac{2\omega R}{c}\,,
\]
where $\omega$, $R$ and $c$ stand for the angular speed and radius of the
central mass, and the speed of light in vacuo. Wilson and Blome (2008)
delivered a lecture in Montreal, and called the attention to the fact that the
most trusted cause for both this anomaly, and the Pioneers, would be
``rotational dynamics.'' I had, by that time, published my results on the
Pioneers Anomaly, through the rotation of the Universe (Berman, 2007). Now, I
shall address the three anomalies.

The first Pioneers Anomaly is the deceleration of about $-9.10^{-8}%
\,\text{cm}.s^{-2}$ suffered by NASA space-probes traveling towards outer
space (Anderson \textit{et al.}, 2002). It has no acceptable explanation
within local Physics, that might also solve the other two anomalies. But, if
we resort to Cosmology, it could be explained by the rotation of the Universe.
Be cautious, because there is no center or axis of rotation. We are speaking
either of a Machian or a General Relativistic cosmological vorticity. It could
apply to each observed point in the Universe, observed by any observer.
Another explanation, would be that our Universe obeys a variable speed of
light Relativistic Cosmology, without vorticities. However, we have shown
elsewhere, that both models are equivalent.

Ni (2008; 2009), has reported observations on a possible rotation of the
polarization of the cosmic background radiation, around 0.1 radians. As such
radiation was originated at the inception of the Universe, we tried to
estimate a possible angular speed or vorticity, by dividing 0.1 radians by the
age of the Universe, obtaining about 10$^{-19}\,\text{rad}.s^{-1}$. Chechin
(2010), and Su and Chu (2009) results are in concordance.

The numerical result is very close to the theoretical estimate, by Berman
(2007),
\[
\omega=\pm c/R=3.10^{-18}\,\text{rad}.s^{-1}, \eqno(8.2.1)
\]
where $c$, $R$ represent the speed of light in vacuum, and the radius of the
causally related Universe.

If we calculate the centripetal acceleration corresponding to the above
angular speed (8.2.1), we find, for the present Universe, with $R\approx
10^{28}\,\text{cm}$ and $c\simeq3.10^{10}\,\text{cm}./s$,
\[
a_{cp}=-\omega^{2}R\cong-9.10^{-8}\,\text{cm}/s^{2}. \eqno(8.2.2)
\]

Our model has been given a General Relativistic cosmological equivalent
treatment (Berman and Gomide, 2010, 2011), with the same results (8.2.1) and
(8.2.2). This value matches the observed experimentally deceleration of the
NASA Pioneers' space-probes. Now, Hubble's constant, in the authoritative
Weinberg's book (Weinberg, 2008), is quoted as $H_{0}=3.10^{-18}s^{-1}$. We
are, thus, tempted to write,
\[
\omega\simeq\pm H. \eqno(8.2.3)
\]

The key result for all these subjects, is that hyperbolic motion, extends
towards infinity, and, thus, qualify for cosmological alternatives, and
boundary conditions. The fly-bys and the Pioneers are in hyperbolic
trajectories, when the anomalies appear, so that Cosmology needs to be invoked.

If we take an imperfect fluid, the Raychaudhuri's equation yields a vorticity
term. But may also, with a non-diagonal metric like Kerr's, introduce
rotations. The most general kind of rotating Universe, that resembles the
Robertson--Walker's one, with perfect fluids, would be a generalized
Robertson--Walker's metric containing a metric temporal coefficient that could
vary with time. In terms of the existing theory of Gaussian metrics (Berman,
2008), we may say that such rotating solution implies that the whole
tri-space, rotates relative to the temporal axis, which is orthogonal to the tri-space.

We, now, shall follow an idea by Godlowski \textit{et al.} (2004), and supply
another General Relativistic model of an expanding and rotating Universe.
Their idea, is that the homogeneous and isotropic models, may still rotate
relative to the local gyroscope, by means of a simple replacement, in the
Friedman--Robertson--Walker's equations, of the kinetic term, by the addition
of a rotational kinetic one (Berman, 2011).

\section{\bigskip}

\bigskip{\LARGE 3. THE GODLOWSKI EQUATIONS}

Consider the flat Robertson--Walker's metric,
\[
ds^{2}=dt^{2}-R^{2}(t)d\sigma^{2}%
\]
Einstein's field equations for a perfect fluid with perfect gas equation of
state, and Robertson--Walker's metric has two. The first, is an energy-density
equation; the second is a definition of cosmic pressure, which can be
substituted by energy momentum conservation. But, upon writing the $\dot
{R}^{2}$ term, we shall add an extra rotational term, namely $(\omega R)^{2}$,
in order to account for rotation. If we keep (8.2.3), the effective Hubble's
parameter, becomes,
\[
H^{2}=\frac{\dot{R}^{2}}{R^{2}}\rightarrow\frac{\dot{R}^{2}}{R^{2}}+
\frac{(\omega R)^{2}}{R^{2}}\approx H^{2}+H^{2}\approx2H^{2}%
\]
and, the field equations become, for a flat Universe,
\[
6H^{2}=\kappa\rho+\Lambda\eqno(8.3.1)
\]
with,
\[
p=\beta\rho\eqno(8.3.2)
\]
and,
\[
\dot{\rho}=-3\sqrt{2}H\rho(1+\beta). \eqno(8.3.3)
\]

The ten field equations reduce to (8.3.1) and (8.3.3), through the standard
Robertson--Walker's metric. As Berman(2011) has pointed out, the Solar system
localized Physics would not be altered by such rotation. First of all, our
rotation is not Godel's. Second, we have argued that hyperbolic motions, which
extend to infinity, and, thus, transcend the local Physical picture, ARE
INDEED, affected, but closed localized orbits, are NOT. Schwarzschild's
metric, does the job without any Hubble's parameter being introduced in the
Solar system. Cosmology has its own rules, and its own observers---co-movers.
Local gravitation, even a General Relativistic one, has other rules and other
observers. The reader should remember that at most, this is an unresolved
issue, for the time being, and one should not adopt radical views.

The usual solution, with Berman's deceleration parameter models, render
(Berman, 1983; Berman and Gomide, 1986),
\begin{align}
R  &  =(mDt)^{1/m},\tag{8.3.4}\\
H  &  =(mt)^{-1},\tag{8.3.5}\\
\ddot{R}  &  =-qH^{2}R=-(m-1)H^{2}R. \tag{8.3.6}%
\end{align}

Notice that we may have a negative deceleration parameter, implying that the
Universe accelerates, probably due to a positive cosmological ``constant.''
But, nevertheless, it is subjected to a negative rotational deceleration, a
kind of centripetal one, that acts on each observed point of the Universe,
relative to each observer, given by relation (8.2.2), so that,
\[
\ddot{R}=-qH^{2}R=qa_{cp}. \eqno(8.3.7)
\]

We now supply the necessary relations among the constants, so that these
equations be observed, namely,
\begin{align}
m  &  =\frac{3}{2}\,\sqrt{2}\,(1+\beta)= \pm\frac{\sqrt{6}}{\sqrt{\kappa
\rho_{0}+\Lambda_{0}}}\,,\tag{8.3.3}\\
\rho &  =\rho_{0}t^{-2},\tag{8.3.9}\\
\Lambda &  =\Lambda_{0}t^{-2}. \tag{8.3.10}%
\end{align}

\section{The Second Pioneers Anomaly}

The angular acceleration of the Universe, taking a positive angular speed, is
given by,
\[
\alpha_{u}=\dot{\omega}=-cH/R=-c^{2}/R^{2}. \eqno(8.4.1)
\]

The spins of the Pioneers were telemetered. And, as a surprise, show that the
on-board measurements yield a decreasing angular speed when the space-probes
were not disturbed. Turyshev and Toth (2010) published graphs (Figures 2.16
and 2.17 in their paper) from which it is clear that there is an angular
deceleration of about 0.1 RPM per three years, or,
\[
\alpha\approx-1.2\times10^{-10}\,\text{rad}/s^{2}. \eqno(8.4.2)
\]

As the diameter of the space-probes is about 10 meters, the linear
acceleration is practically the Pioneers anomalous deceleration value, in this
case, $-6.10^{-8}\,\text{cm}.s^{-2}$. The present solution of the second
anomaly, confirms our first anomaly explanation.

I have elsewhere pointed out that we are in face of an angular acceleration
frame-dragging field, for it is our result (8.4.1). For the Universe, that
causes the result (8.4.2), through the general formula,
\[
\alpha=-\frac{cH}{l}\,, \eqno(8.4.3)
\]
where $l$ is the linear magnitude of the localized body suffering the angular
acceleration frame-dragging.

\section{The Solution of the Fly-By Anomaly}

Consider a two-body problem relative to an inertial system. The same argument
by Godlowski \textit{et al.} (2004), makes us consider that the additional
speed, measured at infinity, relative to the total speed, measured at
infinity, is proportional to twice the tangential speed of the earth,
$w_{e}R_{e}$, divided by the total speed $V+wR\approx c$ taken care of the
Universe angular speed.

This is because we may write,
\[
\frac{\Delta V}{V}=\frac{V+\omega_{e}R_{e}-(V-\omega_{e}R_{e})}{c}=
\frac{2\omega_{e}R_{e}}{c}\approx3\times10^{-6}.
\]

The trick, is that infinity in a rotating Universe, like ours, has a more
precise meaning, when Rotational Cosmology plays the game, because of the
hyperbolic trajectory.

Now, returning to the Machian discussion. To Mach, the definition of
distinguished reference systems was directly related to the dynamical status
of the whole Cosmos. Einstein, on the other hand, was led to define the
inertial systems as those where the laws of nature could be expressed in the
simplest form. But Einstein, after deriving his field equations, as a local
gravitational theory, searched for boundary conditions that could fulfill
Machian theory. If the Universe rotates, as we have hinted, we can explain the
NASA anomalies, we see that General Relativity field equations are correct,
and the Machian dream has come true. We conclude that a rotating and expanding
Universe makes both Einstein and Mach theories correct. I dare to say that
this is probably the unique solution for the Einstein--Machian Program.

\section{\bigskip}

\bigskip{\LARGE 4. THERE ARE NO EXTERNAL AGENTS FOR THE UNIVERSE }\bigskip

The Universe Can Have no External Agents: Improving Einstein's Program

If the Universe has no external \textquotedblleft cause\textquotedblright\ and
if it rotates, it would be interesting to calculate that its spin is constant
in time. In a prior paper (Berman 2011a), it demonstrated the possiblity that
the so-called Einstein's Machian Program could have been finally completed,
through the possible rotation of the Universe; and, after reviewing an earlier
paper (Berman, 2011) published by \textit{Astrophysics and Space Science, }
whereby the Godlowski method of introducing rotation in Cosmology was
presented, then we solved the three NASA anomalies: the linear deceleration of
the Pioneers in outer space, the spin-down of the same space-craft, and the
fly-by anomaly that accompanies gravity assists, with a surplus of kinetic
energy in the end. According to Ni (2011), the ultra-precise Gravity Probe B
experiment analysis, leaves open the cosmic polarization rotation (which may
be due to a Universal rotational state), and the limit of angular speeds
attained by this experiment can be checked from the abstract of his paper to
be around $10^{-17}s^{-1}$. Earlier, Ni (2008; 2009) had estimated the
rotation in 0.1 radians, and we had divided by the age of the
Universe,nfinding a possible angular speed around $10^{-19}s^{-1}$. A recent
paper by Sidharth (2010), placed the angular speed around my (Berman, 2007)
own prediction, $3.10^{-18}\,\text{rad}.s^{-1}$.

Though Einstein originally recognized Machian ideas as important, Barbour
(1990) describes that Einstein concentrated first, into the construction of a
local gravitational theory, delaying consideration on the relativity of
motion, to a future global approach. According to Einstein one should not
speak, in a gravitational theory, on absolute accelerations of a coordinate
system, as much as, in the Special Relativity Theory; one could not work with
absolute speeds of an observer. This apparent betrayal of Machian ideas, was
necessary in order to create the field equations of General Relativity. Mach,
on the other hand, placed the distinguished accelerated reference frame,
within the distant stars, i.e., the local distinguished reference frames,
could be identified by looking at the Cosmos as a whole.

Godlowski (2011) has reviewed the universal rotational evidence. Fine-tuning
arguments can also be invoked in favor of such rotation. Gamow (1946)
considered that a rotation and expansion of the Universe could have the same
physical origin, and we equated the angular speed with the Hubble's parameter.
Chechin (2010) finds the same result, by other token. The present author,
thinks the time is ripe now to reconsider the role of the distant stars
reference frame as a paradigm of accelerations, and introducing the Universal
rotation as proposed by Berman (2007). The angular speed $\omega\cong\frac
{c}{R}\,$ was adopted, because it carries a peculiar rotational state of the
Universe, with deceleration $a_{cp}=-\omega^{2}R$ which coincides with the
Pioneers anomalous deceleration already described. In 2010, Berman and Gomide
(2010) succeeded in making a full General Relativistic model of the rotating
and expanding Machian Universe that was further generalized by Berman and
Gomide (2011a) into a full class of models, thus explaining the three NASA anomalies.

So Einstein's Machian program now has the prototype of a distant stars
absolute centripetal de-acceleration -- it is the Pioneers first anomaly in
action, as we shall see. For the sake of completeness, we mention the three
NASA anomalies, solved by the rotation of the Universe, in the next Section.
Then, we review the theoretical framework, and show that in this model, we may
have a constant Universal spin, the sort of thing that is expected for a
physical system without external causes.

\section{}

Upon considering that Planck's constant renders the spin of Planck's Universe,
Berman (2007; 2010) expected that the Universal Spin could be calculated by,
\[
L=MRc=(4/3)\pi R^{4}\rho c. \eqno(8.7.1)
\]
When we plug (8.3.9), we are left with, imposing constant $L$,
\[
R\propto t^{1/2}, \eqno(8.7.2)
\]
\[
m=2.
\]

Thus, from (8.3.8), we find the equation of state with a negative cosmic
pressure, as expected from the recent Supernovae observations
interpretations,
\[
p=-0.07\rho. \eqno(8.7.3)
\]

\section{\bigskip{\protect\LARGE 5. CONCLUDING REMARKS}}

We have seen that General Relativity accounts for a possible rotating
Universe, and the amount of the deceleration coincides with the P.A. We have
shown that even variable speed of light theory or Machian semi-relativistic
theories also led to the P.A. There is a point that needs clarification.
According to Raychaudhuri's equation, if we consider a non-shearing case, a
non-accelerated system would be described by the equation, adapted to
Robertson--Walker's original metric,
\[
6\ddot{R}=-\kappa\left(  \rho+3p-2\,\frac{\Lambda}{\kappa}\right)
R+4\omega^{2}R, \eqno(8.8.1)
\]
while, in the Generalized Robertson--Walker's metric,
\[
6\ddot{R}=-g_{00}\kappa\left(  \rho+3p-2\frac{\Lambda}{\kappa}\right)  R-
3g_{00}\dot{R}\dot{g}^{00}%
\]
or,
\[
6\ddot{R}=-g_{00}\kappa\left(  \rho+3p-2\frac{\Lambda}{\kappa}\right)  R+
6\dot{R}\omega, \eqno(8.8.2)
\]
but for the Generalized Robertson--Walker's metric, there are two different
solutions, and we would take the negative angular speed solution, in order to
account for a left-handed Universe. In our semi-relativistic treatment of
Chapter 3, what really was needed was a solution for $L^{2}$, so that one
could choose, if necessary in order to coincide with Chapter 4, a negative
angular speed. In Chapter 4, angular speed can also be chosen with a negative
sign; this can be done, for all that matters is the centripetal acceleration,
which depends on the square of $\omega$. However, we must remember that a
positive $\omega=c/R$ also rotates the Universe with the Pioneers deceleration.

We conclude that the Raychaudhuri's vorticity is NOT what we call here the
angular speed of the rotation of the Universe. What we have shown is the
rotation of the entire spatial Universe around the orthogonal time-axis.

By increasing sophistication, we may develop scientific theories which could
\textit{a priori} cover almost any possible characteristic of the Universe
that could actually be observed. By the same token, the reader should remember
that current theoretical cosmological models may easily be turned down by
future astronomical observations. However, at the same time, scientists would
come with ``many'' others, that could be adapted to ``new'' astronomical data,
which on its own, could afterwards go also to the \textit{oblivium}. The
Pioneers Anomaly seems to obey the known laws of Physics, if the Universe
rotates, or the speed of light is variable according to our model. The
secondary anomaly, the spinning down of the spacecraft, received an
explanation as due to the rotation of the Universe. The left-hand of creation
is also accounted by equation (8.8.2) with a negative angular speed. This
introduces a partial decelerating contribution, but if lambda is large enough;
and positive, in order to produce a larger acceleration, $\ddot{R}\geq0,$ and
the Pioneers anomalies will still be there.

\bigskip{\Large Acknowledgements}

I recognize the inspiration from Miss Solange Lima Kaczyk.

{\Large References }

\bigskip

[1] - \bigskip Adler, R.J.; Bazin, M.; Schiffer, M. (1975) -
\ \textit{Introduction to General Relativity, }2$^{nd}$ Edition, McGraw-Hill,
New York.

[2] - Anderson, J.D. et al. (2002)-\textit{ "Study of the anomalous
acceleration of Pioneer 10 and 11".-} Phys. Rev. D 65, 082004.

[3] - \bigskip Anderson, J.D. et al. (2008) - PRL100,091102.

[4] - Berman, M. S. (1981, unpublished) - M.Sc. thesis, Instituto
Tecnol\'{o}gico de Aeron\'{a}utica, S\~{a}o Jos\'{e} dos Campos,
Brazil.Available online, through the federal government site
\ www.sophia.bibl.ita.br/biblioteca/index.html (supply author%
\'{}%
s surname and keyword may be "pseudotensor"or "Einstein").

[5] - \bigskip Berman, M.S. (1983) - \textit{Special Law of Variation for
Hubble%
\'{}%
s Parameter ,}Nuovo Cimento \textbf{74B, }182-186.

[6] - Berman, M.S. (2007) - \textit{Introduction to General Relativistic and
Scalar-Tensor Cosmologies, }Nova Science Publisher, New York. (see Section 7.12)

[7] - Berman, M.S. (2007b) - \textit{The Pioneer Anomaly and a Machian
Universe} - Astrophysics and Space Science, \textbf{312}, 275. Los Alamos
Archives, http://arxiv.org/abs/physics/0606117.

[8] - Berman, M. S. (2008a) - \textit{A General Relativistic Rotating
Evolutionary Universe, }Astrophysics and Space Science, \textbf{314, }319-321.

[9] - Berman, M. S. (2008b) - \textit{A General Relativistic Rotating
Evolutionary Universe - Part II, }Astrophysics and Space Science, \textbf{315,
}367-369.

[10] - \bigskip Berman, M.S. (2011) - \textit{The Two Pioneers Anomalies and
Universal Rotation}, Astrophysics and Space Science, \textbf{336}, 337-339.
DOI 10.1007/s10509-011-0825-4 -

[11] - Berman, M.S. (2011a) - \textit{General Relativity with Variable Speed
of Light and Pioneers Anomaly}, Astrophys. and Space Science, \textbf{336},
327-329. DOI 10.1007/s 10509-011-0839-y

[12] - \bigskip\bigskip Berman, Marcelo Samuel (2012) - \textit{General
Relativity and the Pioneers Anomaly, }Nova Science Publishers, New York.

[13] - \bigskip Berman, Marcelo Samuel (2012a) - \textit{Realization of
Einstein's Machian Program, }\bigskip Nova Science Publishers, New York.

[14] - Berman, M. S. (2012b) - \textit{Realization of Einstein%
\'{}%
s Machian Program: the Pioneers and Fly-By Anomalies}, Astrophys. and Space
Science., \textbf{337}, 477-481, DOI 10.1007/s10509-011-0838-z

[15] - Berman, M.S.; Costa, N.C.A. da (2012) - \textit{On the Stability of our
Universe}, in 2010, Cornell University Library, http://arxiv.org/abs/1012.4160
. Updated in this issue of Journal of Modern Physics (2012).

[16] - \bigskip Berman, M.S.; Gomide, F.M. (1988) - \textit{Cosmological
Models with Constant Deceleration Parameter - }GRG \ \textbf{20,}191-198.

[17] - \bigskip\bigskip Berman, M.S.; Gomide, F.M. (2012) -
\textit{Relativistic Cosmology and the Pioneers Anomaly. }See Los Alamos
Archives (2011) http://arxiv.org/abs/1106.5388 . Published in Journal of
Modern Physics, September, Special Issue on Gravitation, Astrophysics and
Cosmology (this issue).

[18] - Berman, Marcelo Samuel; Gomide, Fernando de Mello (2012a) - \textit{On
the rotation of the Zero-energy Expanding Universe, }in Chapter 12 of
\ \textit{The Big-Bang-Theory, Assumptions and Problems, }ed. by Jason R.
O'Connell and Alice L. Hale, Nova Science Publishers, New York.

[19] - \bigskip Berman, M.S.; Gomide, F.M. (2012b) - \textit{General
Relativistic Treatment of the Pioneers Anomaly, }Journal of Modern Physics,
Special September 2012 Issue on Gravitation, Astrophysics and Cosmology (this
issue). Los Alamos Archives arXiv:1011.4627 (2010).

[20] - \bigskip Chechin, L.M. (2010) - Astron. Rep.54,719.

[21] - \bigskip Godlowski, W.et al. (2004) - Los Alamos Archives,
\ arxiv:astro-ph/0404329 .

[22] - \bigskip Francisco, F. et al. (2011) - Los Alamos Archives,
\ arXiv:1103.5222 .

[23] - \bigskip Cuesta, H.J.M. (2011) - Los Alamos Archives, arXiv:1105.2832 .

[24] - \bigskip Gomide, F.M.; Uehara, M. (1981) - Astronomy and Astrophysics,
\textbf{95}, 362.

[25] - \bigskip\bigskip L\"{a}mmerzahl, C. et al.(2008) - in
\textit{Lasers,Clocks, and dragg-free Control, }ed. by Dittus,Lammerzahl and
Turyshev,Springer, Heidelberg.

[26] - \bigskip MTW - Misner, C.W.; Thorne, K.S.; Wheeler, J.A. (1973)\ -
\textit{Gravitation}, Freeman, San Franscisco.

[27] - \bigskip\bigskip Ni, Wei-Tou(2008) - Progress Theor.
Phys.Suppl.\textbf{172,}49-60.

[28] - \bigskip Ni, Wei-Tou (2009) - Int.Journ.Modern Phys.\textbf{A24,}3493-3500.

[29] - \bigskip Rievers, B.; Lammerzahl, C. (2011) - Los Alamos Archives,
arXiv:1104.3985 .

[30] - Sabbata, de V.; Gasperini, M. (1979) - \textit{Lettere al Nuovo
Cimento, }\textbf{25}, 489.

[31] - So, L.L.; Vargas, T. (2006) - Los Alamos Archives, gr-qc/0611012 .

[32] - Su,S.C.; Chu,M.C.(2009)-Ap.J.703,354

[33] - Turyshev, S.G.;Toth,V.T. (2010) - \textit{The Pioneer Anomaly, }Los
alamos Archives, arXiv:1001.3686

[34] - Weinberg, S.\ (1972) - \textit{Gravitation and Cosmology, }Wiley. New
York.\bigskip

[35] - Weinberg, S.\ (2008) - \textit{Cosmology, }OUP. Oxford.\bigskip

[36] - Wilson, T.L.; Blome, H.-J.(2008)-arXiv:0508.4067

[37] - \bigskip Godlowski, Wlodzimierz (2011) - \textit{Global and Local
Effects of Rotation: Observational Aspects , }Los Alamos Archives,
\ arxiv:1103.5786. International Journal of Modern Physics \textbf{D 20}, 1643.

[38] - Chechin, L. M. (2012) - \textit{Cosmic Structure Formation after the
Big Bang} , Chapter in \textit{The Big Bang: Theory, Assumptions and Problems,
}ed. by Jason R. O'Connell and Alice L. Hale, Nova Science Publishers, New
York. pp.103-172 .

[39] - Raychaudhuri, A.K. (1975) - \textit{Theoretical Cosmology}, Clarendon
Press, Oxford.

[40] - Raychaudhuri, A.K.; Banerji, S.; Banerjee, A. (1992) - \textit{General
Relativity, Astrophysics and Cosmology, }Springer, New York.

\bigskip\lbrack41] - Kerr, R.P. (1963) - \textit{Physical Review Letters,
}\textbf{11}, 237.

[42] - \bigskip Boyer, R.H.; Lindquist, R.W (1967) - \textit{Journal of
Mathematical Physics}, \textbf{8, } 265.

[43] - \bigskip Berman, Marcelo Samuel (2012) - \textit{A\ Primer in Black
Holes, Mach's Principle and Gravitational Energy, }\bigskip Nova Science
Publishers, New York.
\end{document}